\documentstyle[12pt]{article}

\hoffset = - 0.5 truecm
\voffset = - 0.3 truecm
\begin{flushright}
LPTHE-Orsay 94/54\\
hep-ph/9406334
\end{flushright}
\def\beq{\begin{equation}}
\def\eeq{\end{equation}}
\def\noi{\noindent}

\begin{document}
\begin{center}
{\bf ${\bf B}$ TO LIGHT MESON FORM FACTORS} \\
\vskip 5 mm
{\bf R. Aleksan} \\
\vskip 2 mm
Centre d'Etudes Nucl\'eaires de Saclay, DPhPE, 91191 Gif-sur-Yvette, France \\
\vskip 3 mm
{\bf A. Le Yaouanc, L. Oliver, O. P\`ene and J.-C. Raynal} \\
\vskip 2 mm
Laboratoire de Physique Th\'eorique et Hautes Energies\footnote{Laboratoire
associ\'e au Centre
National de la Recherche Scientifique - URA63 - leyaouan@qcd.th.u-psud.fr - fax
33 1 69 41 95
51} \\ Universit\'e de Paris XI, b\^atiment 211, 91405 Orsay Cedex, France \\
\vskip 5 mm
Contribution to ``Beauty 94'' Workshop, April 94, Mont St Michel, France\\
{\it presented by A. Le Yaouanc} \\

\vskip 7 mm
{\bf Abstract :}
\end{center}
The heavy to light form factors in $B$ decays are discussed. Critical
discussion of
theoretical approaches is made with special emphasis on their failure to
describe the $B \to
K^{(\ast )}\psi$ data.

\vskip 1 truecm
\baselineskip 20pt
\noi {\bf I. Motivations} \par
\vskip 5 mm
There is at present no direct experimental data on $B$ to light form factors.
On the other hand, in
addition to its own interest as a test of our understanding of hadron physics,
knowledge of these
form factors is needed for a series of purposes~: \par
i) Determination of the $V_{bu}$ CKM parameter. \par In view of the
difficulties of understanding the
endpoint spectrum in inclusive $b \to u$ decays, especially because of the
importance of
non-perturbative corrections in this region, it seems useful to complement the
standard inclusive
method by the study of exclusive decays. \par ii) Calculation of non-leptonic
decays through WBS
factorization principle. \par
This principle allows to relate the knowledge of non-leptonic decays to the one
of some transition
form factors and annihilation constants. In the non-leptonic decays of class II
like $B \to
K^{(\ast)} \psi$, the needed transition form factors correspond to a $b \to s$
transition. In
class III like $B^- \to D^0 \rho^-$, there is a $b \to u$ transition
contribution. In
fact, most of non-leptonic decays considered for the study of CP violation
involve $B$ to light form
factors. \par It is therefore important to evaluate and to improve the
\underbar{theoretical}
knowledge of these form factors.
\vskip 5 mm

\noi {\bf II. Theoretical estimates : a first discussion} \par

\vskip 5 mm
In spite of the role played by $B$ to-light form factors in several rather
crucial issues, one must
be aware that the theoretical situation is not very good as regards the
estimation of these
quantities. To give a taste of the theoretical situation, it is first advisable
to give the range
of variation of the best determined form factor $f_+$ ($B$ to $\pi$) within a
spectrum of current
approaches~:

$$f_+ = 0.09 \quad \hbox{to} \quad 0.4 \pm 0.1$$

\noindent where the first one corresponds to the very well-known model of Isgur
et al (``GISW''
model) [1] and the second to two-point sum rules [2]. Of course, this is not to
say that all
these estimates are to be considered on equal footing. Precisely, we want to
make a critical
survey. We shall first show that we have not very trustable predictions because
of theoretical
weaknesses or sources of uncertainties inherent to most theoretical approaches,
or even because of
definite experimental failures of popular quark models already in the $D$'s
(this section). In a
second step this criticism will be extended by showing that strong
discrepancies with experiment are
also present in these approaches at the $B$ level and already visible (section
3) [3]. \par

\vskip 5 mm
\noi {\bf 2.2 Fundamental methods}
\vskip 5 mm
By this, we mean methods which are closely based on QCD, as opposed to more
phenomenological
approaches like the quark model. A priori, since the problem at hand seems a
highly relativistic one
(strong binding for light quarks, large possible transfer $\vec{q}$), one would
be tempted to rely
essentially on such methods. However, they are finally rather disappointing at
present for $B$
mesons, in contrast to the case of $D$ mesons. \par

$\bullet$ \underbar{Numerical lattice QCD} \par This is the most
fundamental approach but it is very disappointing in the problem at hand.
Indeed, one faces two main
difficulties~: \par 1) The $B$ mesons are too heavy to be directly put on the
lattice. \par

2) large $\vec{q}\ $'s which necessarily appear in $B$ decays (for instance $t
= 0$ correspond to
$\vec{q} \sim m_B/2$ in $B \to \pi$) are affected by very large errors. \par

Therefore one has to proceed [4] to two extrapolations~: \par

1) Extrapolation in mass from the region where one can actually propagate heavy
mesons, i.e. the
$D$ region, which is however rather far from the $B$. But $B$ is indeed not so
far from $D$ in $1/m$,
which is the relevant parameter. \par

2) Extrapolation in $\vec{q}$ or $t$ from the region where one can actually
have accuracy, i.e. the
region of small $\vec{q}$ - for which we have however few points, and which is
far from $t = 0$ in
the $B$. \par

The first extrapolation seems to be doable with enough statistics in a near
future. The second one
is rather difficult and the results around $t \sim 0$ depend very strongly on
the assumed
extrapolation formula in $t$. The most convincing result is the finding of a
dependence on the heavy
quark mass which is smoothed with respect to its asymptotic form. \par

$\bullet$ \underbar{QCD sum rules} \par A priori, QCD sum rules do not have
such severe limitations
and can treat directly the $B$'s. It appears however that problems still arise
in the limit $m_Q \to
\infty$. For transition form factors, double Laplace sum rules [5] have certain
diseases, which seem
to be cured by Narison's hybrid sum rules [6]. \par

However, we still observe problems in the latter approach. First there are very
large errors in $B
\to \pi$. Second, there is simply no stability for $B \to \rho$ and also very
large errors. These
large errors are not apparent in final results where the hybrid sum rules and
double
Laplace sum rules results are amalgamated ; since they have in fact a small
overlap, one ends with
very small errors. The $t$ dependence seems also rather surprising using both
methods. For instance
$f_+$ seems pole-like according to P. Ball [5], while she predicts a decreasing
$A_1$ from
$t = 0$ up to $15 \ {\rm GeV}^2$. \par

On the whole, the situation seems to us rather uncertain. \par

\vskip 5 mm
\noi {\bf 2.3 Quark models}
\vskip 5 mm
Actually, facing this rather disappointing situation of fundamental methods,
one tends to regard
with more consideration the old phenomenological approach of quark models,
especially since it
yields definite numbers for all values of mass and transfer and for excited
states as well. This
motivation however is perhaps not excellent by itself. Rather, one should
emphasize that this large
covering  relies in addition on a solid physical ground~: the one of bound
state physics, which
give an essential conceptual insight not provided by fundamental methods. \par

It is too loose, however, to speak about quark models in general. Indeed, the
most popular models
of form factors, GISW [1] and WBS [7], [8], are not really satisfactory. And
the good approach is not
to be found in nonrelativistic models which have been mostly developed
formerly, but rather in
relativistic approaches, which, although pursued since a long time [10] have
escaped attention, and
moreover have not yet been elaborated into a complete model accounting for all
important
relativistic effects. \par

$\bullet$ The GISW model is a basically non-relativistic one, but for a crucial
adhoc adjustment of
$t$ dependence through

\[\vec{q}\ ^2(\hbox{rest frame}) \longrightarrow {1 \over \kappa^2} \quad
m_f/m_i \quad (t_{max}-t)\]

\noi there is here no theoretical deduction, but again a rather arbitrary
extrapolation from the
truly non-relativistic point $t_{max}$ $(\vec{q} = 0$) to the highly
relativistic $t = 0$. One must
note that distinction should be made between the GISW quark model for
spectroscopy which may be
quite reasonable, and the one for form factors, which relies on a complete
alteration of the
parameters of the latter through the fudge factor $\kappa^2 \sim 0.5$. \par

$\bullet$ The WBS models are quite different in spirit. The real quark model
calculation is done
only at $t = 0$ (rather than at $t_{max}$). It claims to be relativistic. It is
combined with
assumptions concerning the $t$ dependence starting from $t = 0$, of power-like
type. \par

One must stress, since it is not always clearly recognized, that there are two
WBS models
corresponding to two such different assumptions for heavy to light
transitions~: the older one [7]
assumes VMD behavior for all form factors, the new one [8] mixes a pole-like
behavior for certain
form factors and a dipole one for others. \par

One must also be aware that these latter assumptions are something quite
external to what must be
properly termed as the WBS {\it quark} model, which is a $P = \infty$
approach~; this one is used
only at $t = 0$, while in principle it could predict form factors at any $t$.
\par

Against these popular models, there are both experimental and theoretical
objections~: \par

1) In the $D$ decays, the $D \to K^{\ast}\ell \nu/K \ell \nu$ ratio is
predicted much too large and
$\Gamma_L/\Gamma_{tot}$ too small [9]. These facts can be interpreted by saying
that $A_1$
and $A_2/A_1$ are too large. \par

2) These models do not reproduce the Isgur-Wise scaling in the limit $m_Q$,
$m_q \to \infty$. \par
The answers to these difficulties are partly found in the relativistic
treatment of spin, not
present in these models~: \par

$\alpha$) The correct \underbar{relativistic boost of spin}, as shown by us
[10], automatically
introduces the wanted ratios of form factors required by Isgur-Wise scaling. In
particular the
ratios of $f_+$, $V$, $A_2$ against $A_1$ increase with $t$. It is also found
that this behavior in
$t$ is maintained for heavy to light transitions, with violation of scaling
mostly independent of
$t$. \par

$\beta$) The effect of the small component of Dirac spinors due to the large
internal velocities
reduces $A_1$ and therefore $K^{\ast}/K$. \par

In addition, the GISW model is missing another important relativistic effect~:
the Lorentz spatial
contraction of wave functions which would smooth the gaussian fall-off of form
factors at large
$\vec{q}$. This is especially crucial for $B$ decays near $t = 0$ and for the
$B \to \pi$
transition, which explains the very small value obtained in this case. It seems
that, on the
contrary, this effect is included in the WBS quark model. \par

On the whole, it seems that by inclusion of the two effects of the relativistic
center-of-mass
motion~: Lorentz contraction and boost of spin, as well as of the effect of
binding, one can
account for a good part of the problems [11]. \par

It is also interesting to note that the quark model formulae are naturally
presenting a smooth
dependence in the heavy mass $m_Q$ ($D$, $B$, ...) showing some similar
tendencies to what is found
by lattice QCD. This is found already in a non-relativistic calculation. \par

On the other hand, all the quark models fail at present to explain the relative
smallness of
$A_2/A_1$ which is found for $D \to K^{\ast}$ by experiment, and which has been
predicted by
lattice QCD. \par

Finally, it must be stressed that no quark model at present includes all the
above effects. The
best we can offer at present is our weak-binding Orsay quark model [10], [3]
which includes Lorentz
spatial contraction effect and the relativistic boost of spin~: it can claim
only to describe the $t$
dependence. Our conclusion is that $A_1$ is very slowly increasing, while
$f_+$, $V$, $A_2$ should
have pole-like behaviour. At large $\vec{q}\ ^2$ ($t \simeq 0$), this is
happily not very dependent
on the details of the potential.

\vskip 5 mm
\noi{\bf 2.4 Conclusion}
\vskip 5 mm

If we consider the theoretical situation on the overall, we can conclude that
it is far from being
satisfactory. 1) Lattice QCD requires drastic extrapolations. 2) Sum rules
yield very uncertain
results (although this seems to have been recently improved by considering
ratios of form
factors). 3) Quark models, even if relativistic, are limited to relatively weak
binding, which is
especially worrying for the $\pi$.

\vskip 5 mm
\noi{\bf III. Returning to experiment : indications from ${\bf B}$ data}
\vskip 5 mm
Therefore theory cannot make safe predictions, and it cannot be trusted.
Experimental knowledge
cannot be avoided, and is also required for a further elaboration of theory.
\par

The question is now~: what can we learn \underbar{at present} about $B$ to
light transition
form factors, given that the corresponding semi-leptonic transitions are very
weak, that we do
not know $V_{bu}$ better than very roughly - and that finally one of the aims
would be to disentangle
$V_{bu}$ from experiment and $b$ to $u$ mesons form factors~? \par

If one admits, as is strongly suggested by the findings of fundamental methods,
that $SU(3)$ is
not very strongly broken, there is a possibility to circumvent the problem by
looking at $b$ to
strange transition which will not involve  $V_{bu}$ but only standard Cabibbo
angles and $V_{bc}$.
Two examples are~: \par

i) $B \to K^{\ast}\gamma$. Recently measured. Since QCD coefficients seem
trustable, one has
direct access to one form factor at $t = 0$, $f_1(0)$. It is related by Heavy
Quark Symmetry to the
semi-leptonic form factors~: $f_1 \sim 1/2$ $(V + A_1)_{t=0}$. This combination
is however not very
sensitive to the different theoretical estimates, and present experiment is
compatible with most of
them. \par

ii) $B \to K^{(\ast )}\psi$. Assuming the WBS factorization, one has access to
form factors at
$t = m_{\psi}^2$ through two ratios~: $K^{\ast}$ polarization
$\Gamma_L/\Gamma_{tot}$ and
$K^{\ast}/K$ [12] ; the absolute normalization is not accessible since it
depends on the adhoc
WBS factor $a_2$, but can be constrained partly by considering the output value
of $a_2$, which
cannot be too large. \par

$\bullet$ \underbar{The failure of current theoretical approaches in $B \to
K^{(\ast)}\psi$} \par
 Anyway, it is remarkable that the two ratios by themselves already allow a
very
dis\-cri\-mi\-na\-ting test. Most theoretical estimates fail at predicting one
or both of these
ratios, often very badly, as shown by table 1. \par

\begin{center}
\begin{tabular}{|l|c|c|} \hline
& & \\
&$\Gamma_L/\Gamma_{tot}$ & ${\Gamma (B \to K^{\ast} \psi ) \over \Gamma (B \to
K \psi )}$ \\
& & \\ \hline
\enskip Quark models & & \\ \hline
\enskip WBS I & 0.58 & 4.23 \\ \hline
\enskip WBS II & 0.36 & 1.62 \\ \hline
\enskip GISW & 0.06 & 0.95 \\ \hline
\enskip Sum rules & & \\
\enskip P. BALL & 0.36 & 7.6\ \\ \hline
\enskip Lattice QCD data & & \\
\enskip of ABADA et al, & 0.20 $\pm$ 0.20 & 1.8 $\pm$ 1.6 \\
\enskip with a new analysis   & & \\ \cline{2-3}
\enskip by PENE with two & & \\
\enskip different assumptions  &0.42 $\pm$ 0.15 & 3.2 $\pm$ 2.8 \\ \hline
\enskip Experience (CLEO) & 0.78 $\pm$ 0.10 $\pm$ 0.10 & 1.52 $\pm$ 0.43 \\
\hline
\end{tabular}
\end{center}

\begin{center}
Table 1
\end{center}
\vskip 5 mm

One observes that sum rules as well as the popular quark models fail, WBS II
failing less badly
than others. \par

Lattice QCD data would be unconclusive if one was not to add additional
assumptions for the $t$
behavior. Yet, they do not show a clear agreement and are still very uncertain.
\par

The failure sometimes concerns the polarisation ratio $\Gamma_L/\Gamma_{tot}$,
and in other
cases the $K^{\ast}/K$ ratio. Combining the two ratios, it appears that one has
found a rather
discriminating test. \par

$\bullet$ \underbar{How shall we interpret this failure ?} \par

- either WBS factorisation fails in class II. This is a possibility, although
it is not so bad even
for $D$'s. \par

- or experience has some weakness. Indeed, at present, the situation is not so
satisfactory with
the three sensibly different central values of Argus, Cleo and CDF for
$\Gamma_L/\Gamma_{tot}$ while
$K^{\ast}/K$ is only given by CLEO. \par

- or finally, as seems to us the more probable, the blame must be put on
current theoretical
estimates of form factors. Indeed, none of these have a sufficiently solid
basis to be considered as
crucial. The relative agreement of fundamental methods like sum rules in the
case of $D$'s - which
are still poorly measured - does not guarantee that they also work for $B$'s,
while popular quark
models already fail in $D$'s. \par
\pagebreak
$\bullet$ \underbar{Our solution} \par
In our opinion, a combined analysis of $B \to K^{(\ast)} \psi$
and $D \to K^{(\ast )} \ell \nu$ data seems to favor the last interpretation.
Indeed we find a
phenomenological model [3] which is based on an extension of heavy to heavy
Isgur-Wise scaling to
finite masses and to light quarks, with a rescaling of the various form factors
independent of the
masses and of the momentum transfer $t$ to describe the $D \to K^{( \ast )}
\ell \nu$ data, and
which describes rather satisfactorily $B \to K^{(\ast )}\psi$, although with a
too large $\chi^2$.
\par

As regards quark-mass and $t$ dependences, this model is in fact a simplified
version of the
semi-relativistic quark model for form factors developed by the Orsay group,
which shows that the
heavy to heavy Isgur-Wise scaling laws extend partly to heavy to light
transitions. The rescaling of
the form factors in the phenomenological model may be thought as corresponding
theoretically to two
types of effects \par

1) finite mass effect of the spectator quark, included in the quark model  \par

2) binding corrections, not included in the quark model presently. \par

This quark model behaves differently from all popular quark models especially
by its $t$ behaviour.
It also differ from the various versions of the sum rules. It could be tested
in detail in a
$\tau$-charm factory. \par

$\bullet$ \underbar{Improving present tests and finding new ones} \par
Of course it is still too early to conclude, and new experimental inputs are
highly desired. In
the near future, one could envisage the following~: \par a) First of all more
precise measurements of
$D \to K^{(\ast)} \ell \nu$ and $B \to K^{(\ast)} \psi$ would be crucial to
draw firmer conclusions
from the present analysis, and one of the closest goals is the polarisation
data in $B \to K^{\ast}
\psi$. \par

b) The approach to $B$ to light form factors through non leptonic decays could
be extended by
measuring class II decays of the type $D^{0} \pi^0$ ... and by improving
accuracy on class III.
These two types of data would give access to $b$ to $u$ transitions. \par

To be able to use quantitatively the class III data, it must be stressed that
the ratio
$\tau_{B^-}/\tau_{B^0}$ must be known accurately, in order to convert models
into predictions for
branching ratios. Yet, this ratio can be expected to be very close to 1. \par

c) More precise data on $B \to K^{\ast}\gamma$ could increase the presently low
sensitivity to model
variation. \par

d) Baryon decays like $\Lambda_b \to \Lambda \psi$ ... could allow clear tests
of factorisation
since baryon matrix elements are more controllable theoretically in quark
models. \par
\vfill \pagebreak
%\vskip 5 mm
\noindent {\bf Acknowledgments :} \par
I would like finally to thank all the organizers and I. Bigi for having the
pleasure to talk in
this place, so close to the part of Brittany where I have spent many years.
This work was supported in part by the Human Capital
and Mobility Programme, contract CHRX-CT93-0132.\par

\vskip 5 mm
\noindent {\bf References} \par
\vskip 5 mm
\begin{description}
\item{[1] \ } N. Isgur, D. Scora, B. Grinstein and M. B. Wise, Phys. Rev. D39
(1989)
799. \par
\item{[2] \ } C. A. Dominguez and N. Paver, Z. Phys. C41 (1988) 217. \par
\item{[3] \ } R. Aleksan, A. Le Yaouanc, L. Oliver, O. P\`ene and J.-C. Raynal,
preprint LPTHE 94-15.
\par \item{[4] \ } A. Abada et al., Nucl. Phys. 416 (1994) 675. \par
\item{[5] \ } P. Ball, Phys. Rev. D48 (1993) 3190. \par
\item{[6] \ } S. Narison, Phys. Lett. B283 (1992) 384. \par
\item{[7] \ } M. Wirbel, B. Stech and M. Bauer, Z. Phys. C29 (1985) 637. \par
\item{[8] \ } M. Neubert, V. Rieckert, B. Stech and Q. P. Xu, \underbar{Heavy
Flavours}, Editors A.J.
Buras and M. Lindner, World Scientific, Singapore, 1992. \par
\item{[9] \ } N. Isgur, ``Models of semileptonic decays'', Invited talk at the
1989 Heavy Quark
Symposium, Cornell, June, 1989, report UTPT-89-25. \par \item{[10]} A. Le
Yaouanc, L. Oliver, O.
P\`ene, J.-C. Raynal, Gif lectures 91 (electroweak properties of heavy quarks),
tome 1, p. 89 and
references therein.\par  \item{[11]} A. Le Yaouanc and O. P\`ene, Proceedings
of the Marbella Workshop
on tau-charm factory (1993).\par
\item{[12]} A. Freyberger, SLAC Summer Institute 1992, SLAC-report 412, page
415. \par

\end{description}
\end{document}